\documentstyle[preprint,aps]{revtex}
\begin{document}
\title{Transitions to equilibrium state in classical $\phi ^{4}$ lattice}
\author{Xingang Wang$^{1},$ Ying Zhang$^{2}$ and Gang Hu$^{3,1}$\thanks{%
Author for correspondence}}
\address{$^{1}$Department of physics, Beijing Normal University, Beijing\\
100875, China\\
$^{2}$The Abdus Salam International Centre for Theoretical Physics(ICTP),\\
P.O. Box\\
586, 34100 Trieste, Italy \\
$^{3}$China Center for Advanced Science and Technology (CCAST)\\
(World Laboratory), P.O. Box 8730, Beijing 100080, China}
\date{\today}
\maketitle

\begin{abstract}
Statistical behavior of a classical $\phi ^{4}$ Hamiltonian lattice is
investigated from microscopic dynamics. The largest Lyapunov exponent and
entropies are considered for manifesting chaos and equipartition behaviors
of the system. It is found, for the first time, that for any large while
finite system size there exist two critical couplings for the transitions to
equipartitions, and the scaling behaviors of these lower and upper critical
couplings vs the system size is numerically obtained. \newline
\end{abstract}

\newpage

\section{introduction}

The investigation of statistical behavior of Hamiltonian coupled oscillators
from microscopic dynamics has attracted much attention since the original
Fermi-Pasta-Ulam (FPU) problem [1-4]. The important issue in this aspect is
the approach to equilibrium of nonlinear system with many degrees of
freedom. It was found in the FPU study that with sufficiently small energy
level various modes of the system do not go to equipartition state as
conventionally expected for equilibrium [1]. This observation has stimulated
many works studying the transitions from nonequilibrium to equilibrium,
based on numerical simulations of microscopic dynamics [5-9]. Various models
of Hamiltonian system with many degrees of freedom have been considered for
this purpose, among which the $\phi ^{4}$ lattice is one of the most
extensively investigated\cite{sst}.

The model Hamiltonian reads 
\begin{eqnarray}
\widehat{H} &=&%
\mathrel{\mathop{\sum }\limits_{i=1,....N}}%
(\frac{m\widehat{p}_{i}^{2}}{2}+\frac{\mu (\widehat{q}_{i}-\widehat{q}%
_{i+1})^{2}}{2}+\frac{\beta \widehat{q}_{i}^{4}}{4}) \\
\widehat{q}_{N+1} &=&\widehat{q}_{1}\text{ , }\widehat{p}_{N+1}=\widehat{p}%
_{1}  \nonumber
\end{eqnarray}%
With the rescalings $q_{i}=\sqrt{\mu /\beta }\widehat{q}_{i},$ $p_{i}=\sqrt{%
m/\beta }\widehat{p}_{i}$ and $t=\sqrt{m/\mu }\widehat{t}$, we can reduce
Eq. (1)\ to $H=%
\mathrel{\mathop{\sum }\limits_{i=1,....N}}%
(\frac{p_{i}^{2}}{2}+\frac{(q_{i}-q_{i+1})^{2}}{2}+\frac{q_{i}^{4}}{4}).$ So
far most of papers considering the statistical behavior of the lattice $\phi
^{4}$ model have used this form and taken the average initial energy as the
control parameter. In this paper, we use an alternative while equivalent
method to fix the initial average energy $\overline{E}=\frac{1}{N}%
\mathrel{\mathop{\sum }\limits_{i=1,...N}}%
E_{i}$ and vary the coupling strength $\mu $ as the control parameter, i.e.,
we consider the lattice $\phi ^{4}$ Hamiltonian%
\begin{equation}
H=%
\mathrel{\mathop{\sum }\limits_{i=1,....N}}%
(\frac{p_{i}^{2}}{2}+\frac{\mu (q_{i}-q_{i+1})^{2}}{2}+\frac{q_{i}^{4}}{4})
\end{equation}%
which leads to the set of canonical equations%
\begin{eqnarray}
\stackrel{.}{p}_{i} &=&-\frac{\partial H}{\partial q_{i}}=-q_{i}^{3}+\mu
(q_{i+1}+q_{i-1}-2q_{i})  \nonumber \\
\stackrel{.}{q}_{i} &=&\frac{\partial H}{\partial p_{i}}=p_{i}
\end{eqnarray}%
We take $\overline{E}=1.25\times 10^{-3}$ throughout the paper without
lossing any generality.

Before going to the detail of the dynamics of Eqs. (3), we can first briefly
discuss the general feature of the system for different coupling regimes.
For $\mu \rightarrow 0,$ the system consists of practically independent
oscillators, each has one degree of freedom, and thus has nearly periodic
motion. The dynamics of the system may be quasiperiodic and the approach to
equilibrium is not possible. For $\mu \rightarrow \infty ,$ the nonlinear
terms $q_{i}^{4}/4$ become not important in comparison with the coupling.
Thus, different \ modes are practically uncoupled to each other [10-12], and
the system dynamics is quasiperiodic again. Complicated dynamics approaching
equilibrium may be expected for not too small and not too large couplings
(i.e., not too large and not too small energies if we fix $\mu =1$ and use
the average energy as our control parameter). Up to date, all papers
discussing the statistical behaviors of the $\phi ^{4}$ model consider only
small energy (i.e., large coupling $\mu $) regime. To our knowledge, the
behaviors of large energy (or say, small coupling $\mu $) regime has never
been discussed. In this paper we will study the microscopic dynamics of the
system and the related statistical features in the whole coupling range by
paying particular attention to the new area, the small coupling regime.

In the following, we will make a Lyapunov exponent analysis of the system
dynamics in Sec. II; and study the statistical quantities, the entropy and
the probability distribution of system variables in Sec. III. In Sec. IV, we
will consider the scaling behaviors observed in the transitions to
equilibrium, and some conclusions and possible future developments will be
discussed in the last section.

\section{Lyapunov exponent analysis}

Given the average energy $\overline{E}$ and the coupling $\mu ,$ the
dynamics of Eqs. (3) is determined by the initial conditions of the $%
(q_{i},p_{i})$ variables. In this paper we will test and compare two types
of typical initial conditions. First, the initial energy is distributed to a
single $k=1$ mode%
\begin{eqnarray}
q_{i} &=&0\text{ },  \nonumber \\
p_{i} &=&A\sin (2\pi i/N)+\sigma _{i}\text{ },\text{ }|\sigma _{i}|\ll 1%
\text{ , }A=\sqrt{4\overline{E}} \\
i &=&1,2,...,N  \nonumber
\end{eqnarray}%
i.e., 
\begin{eqnarray*}
Q_{k} &=&0\text{ , }k=0,1,2,....N/2 \\
P_{1} &=&A\sqrt{N/2}\text{, }P_{k}=0\text{ , }k=0,2,3,....,N/2 \\
Q_{k}(t) &=&\sqrt{2/N}\stackrel{N}{%
\mathrel{\mathop{\sum }\limits_{i=1}}%
}q_{i}(t)\cos (2\pi ki/N),\text{ }P_{k}(t)=\stackrel{.}{Q}_{k}(t)
\end{eqnarray*}%
This initial condition has been popularly used in studying the statistical
behavior of Hamiltonian coupled oscillators\cite{sst}. Second, the initial
excitation is restricted on a single site%
\begin{eqnarray}
q_{i} &=&0\text{ },\text{ }i=1,2,....,N  \nonumber \\
p_{1} &=&\sqrt{2N\overline{E}}\text{ \ \ and \ }p_{i}=\sigma _{i}\text{ , }%
i=2,3,...N
\end{eqnarray}%
which is considered for the first time, to our knowledge, in this paper.
Here the equipartition problem will be analysed for both mode distribution
and site distribution. For avoiding the strict symmetry restriction
preventing equipartitions we add a small random numbers $\sigma _{i}$ in
Eqs. (4)\ and (5) with $\left| \sigma _{i}\right| \leq 10^{-7}.$

Lyapunov exponent (LE) analysis is a very useful dynamical tool for
revealing the statistical behavior of Hamiltonian systems [10,12,13]. If the
dynamics is quasiperiodic, the largest LE is zero, and the system will never
reach the equal-probability-distribution in the energy surface for $N\geq 2$%
. Because for a Hamiltonian system with $N$ degrees of freedom the
quasiperiodic trajectory can move only in a $N$-dimensional torus, while the
energy surface has $2N-1$ dimensions. Therefore, the positivity of the
largest LE is a strong indication (of course, not a sufficient condition) of
possible equilibrium state.

It is well known that there exist some difficulties for correctly computing
LEs in Eqs. (3)\ when $N$ is large [10,12]. Numerically, LE is calculated
approximately with a very long but still finite time $T$. However, if $N$ is
large, it may happen that the system wonders in a certain $N$-dimensional
torus for an extremely long time and then bursts to a chaotic status. In
Figs. 1(a) and (b) we take $N=64$ and use the $k=1$ mode initial condition
(4),\ and plot the largest LE $\lambda _{1}$ against the average time $T$
for $\mu =6.9\times 10^{-7}$ and $2.5$, respectively$.$ In both case we find
that $\lambda _{1}$ is maintained at zero for long time, and jumps to be
positive at a certain large $T$. If we decrease $\mu $ from $\mu =6.85\times
10^{-7}$ and increase $\mu $ from $\mu =2.9,$ the periods for $\lambda _{1}$
to change from zero to positive increase. It seems that there exist two
critical couplings $\mu _{1}$ and $\mu _{2},$ and zero $\lambda _{1}$ is
observed for $\mu <\mu _{1}$ or $\mu >\mu _{2},$ and chaos is identified
between these two couplings $\mu _{1}<\mu <\mu _{2}.$ Transitions from
quansiperiodicity to chaos occur at both large $(\mu _{2})$ and small $(\mu
_{1})$ critical couplings. For approximately fixing $\mu _{1}$ and $\mu
_{2}, $ we take average time $T_{0}=1\times 10^{7},$ and zero $\lambda _{1}$
is assumed if no burst to positive LE accurs within the period $T_{0}.$

In Fig. 1(c) we plot $\lambda _{1}$ vs $\mu $ of Eqs. (3) at $N=32$ and for
both initial conditions (4) (circles)\ and (5) (triangles). In each curve we
find indeed critical couplings $\mu _{1}(\mu _{1}^{^{\prime }})$ and $\mu
_{2}(\mu _{2}^{^{\prime }})$, and $\lambda _{1}$ is zero for $\mu <\mu _{1}$ 
$(\mu <\mu _{1}^{^{\prime }})$ and $\mu >\mu _{2}$ $(\mu >\mu _{2}^{^{\prime
}}).$ Positive $\lambda _{1}$, indicating chaos, is observed between these
two critical couplings. Another interesting point is that the circle and
triangle lines have practical identical positive $\lambda _{1}$ in the
region $\mu _{1}^{^{\prime }}<\mu <0.25,$ this verifies approach to
equilibrium, which is independent of the initial conditions.

In Fig. 1(d) we do the same as (c)\ by taking larger system $N=64$. For the
initial condition (4), the lower (higher)\ critical coupling $\mu _{1}(\mu
_{2})$ becomes much smaller (much larger) than that in (c). This feature is
intuitively expected since it is familiar to us that for large $N$ one needs
only small interactions between the subsystems to bring the system to
equilibrium states\cite{revisite}. [note, for Eqs. (3) large coupling $\mu $
corresponds to small interactions between different modes]. However, for the
initial condition (5), the critical coupling $\mu _{1}^{^{\prime }}$
increases as the system size increases. This phenomenon, which seems to be
against the intuition, can be easily understood, based on the measue
synchronization (MS) of Hamiltonian systems\cite{MS}. In order to reach the
equilibrium state, all the oscillators of Eqs. (3)\ should have identical
statistical behavior, independing on their initial condition, because they
are identical in microscopic dynamics. The necessary condition realizing
this identity is that the trajectories of all oscillators should be
identical in measure, so-called MS. The critical coupling for the MS between
any two neighbor oscillators depends on their energy difference. Large
coupling is needed for the MS if a pair of neighbor oscillators have large
energy difference. Since for condition (5) the energy difference between the
site $i=1$ and its neighbors increases by enlarging the system size with
fixed average energy [in (5) $E_{1}-E_{2}=N\overline{E}$], the critcal
coupling for MS, which is nothing but $\mu _{1}^{^{\prime }}$, increases as
well for increasing the system size (actually $\mu _{1}^{^{\prime }}\propto 
\sqrt{N}$). This feature is essentially different from the system behavior
with the initial condition (4).

\section{Entropies and probability distributions}

The dynamic behavior of LE is closely related to the statistical behavior of
the system. It is interesting to further investigate the energy distribution
of subsystems, because the equipartition behaviors of energies of different
modes and different sites are some necessary conditions of the equilibrium
state.

The energies for different modes can be computed by%
\begin{equation}
E_{m}(k,t)=\frac{1}{2}P_{k}^{2}(t)+\frac{1}{2}\omega _{k}^{2}Q_{k}^{2}(t)%
\text{, \ }k=0,2,....\frac{N}{2}
\end{equation}%
where $\omega _{k}=2\sin (\frac{k\pi }{N})$. At the same time, the energy
for different sites are represented as%
\begin{equation}
E_{s}(i,t)=\frac{1}{2}p_{i}^{2}(t)+\frac{1}{4}q_{i}^{4}(t)\text{, \ }%
i=1,2,.....N
\end{equation}%
For the long time behavior, we are interested in the averages%
\begin{eqnarray}
\overline{E}_{m}(k,T) &=&\frac{1}{T}\int_{0}^{T}E_{m}(k,t)dt  \nonumber \\
\overline{E}_{s}(i,T) &=&\frac{1}{T}\int_{0}^{T}E_{s}(i,t)dt
\end{eqnarray}%
In actual simulations we take large while finite $T=10^{7}$.

Based on the definition of Eqs. (8), we can investigate the system
statistics by defining entropies\cite{entropy} in the mode space%
\begin{eqnarray}
S_{m} &=&-\stackrel{N/2}{%
\mathrel{\mathop{\sum }\limits_{k=0}}%
}\Omega _{k}\ln \Omega _{k} \\
\Omega _{k} &=&\overline{E}_{m}(k)/\stackrel{N/2}{%
\mathrel{\mathop{\sum }\limits_{k^{^{\prime }}=0}}%
}\overline{E}_{m}(k^{^{\prime }})  \nonumber
\end{eqnarray}%
and in the site space%
\begin{eqnarray}
S_{s} &=&-\stackrel{N}{%
\mathrel{\mathop{\sum }\limits_{i=1}}%
}\Omega _{i}^{^{\prime }}\ln \Omega _{i}^{^{\prime }} \\
\Omega _{i}^{^{\prime }} &=&\overline{E}_{s}(i)/\stackrel{N}{%
\mathrel{\mathop{\sum }\limits_{i^{^{\prime }}=1}}%
}\overline{E}_{s}(i^{^{\prime }})  \nonumber
\end{eqnarray}%
In numerical computation we measure the scaled entropy differences%
\begin{eqnarray}
\eta _{m} &=&(S_{m}(\max )-S_{m})/S_{m}(\max )  \nonumber \\
\eta _{s} &=&(S_{s}(\max )-S_{s})/S_{s}(\max )
\end{eqnarray}%
with $S_{m}(\max )=\ln (N/2+1)$, $S_{s}(\max )=\ln (N).$ In Fig. 2(a) [(b)]
we plot $\eta _{m}$ ($\eta _{s}$) vs $\mu $ from the initial condition (4)\
[the initial condition (5)]. We find that in (a)\ $\eta _{m}$ has
discontinuous jumps at the two critical couplings $\mu _{1}$ and $\mu _{2}$
and it takes very small value for $\mu _{1}<\mu <\mu _{2}$, indicating large 
$S_{m}$ entropy in this coupling regime. In (b) $\eta _{s}$ has a plateau of 
$\eta _{s}\approx 1$ for $\mu <\mu _{1}^{^{\prime }},$ and it jumps down to
about $\eta _{s}<1\times 10^{-6},$ and keeps this small value in the entire
regime $\mu >\mu _{1}^{^{\prime }}.$ It is noticed that the second
transition at $\mu _{2}^{^{\prime }}$ is not manifested by $\eta _{s}$,
because the MS between sites is not affected by this transition.

In order to show that in the regime $\mu _{1}^{^{\prime }}<\mu <1.65$, the
system actually reach the statistical equilibrium state for both initial
conditions, we take $\mu =0.1$ and plot in Figs. 2(c)\ and (d)\ the
probability distributions of $x$ and $p$ variables, respectively, for all
the $64$ particles and for both initial conditions (4)\ and (5). Each
probability distribution curve is drawn with $1\times 10^{6}$ data. From the
figures it is verified that all the sites have identical distributions,
showing perfect equipartion behaviors.

\section{scaling behaviors in the transitions to equilibrium}

In the above numerical analysis we found two transitions of Eqs. (3)\ for
the initial condition (4) from quansiperiodicity to chaos, one at small
coupling $\mu _{1}$, transition to equipartion between different sites in
real space, the other at large coupling $\mu _{2}$, transition to
equipartition between various modes in Fourier space. Both $\mu _{1}$ and $%
\mu _{2}$ can be considerably changed by varying the system size. It is
interesting to investigate the possible scaling behavior between the
critical $\mu $ and the system size $N.$

In Fig. 3(a) we plot both $\ln \mu _{1}$ and $\ln \mu _{2}$ vs $\ln N.$
Between the two curves the system motion is chaotic, where statistical
equipartition behavior is expected. And quasiperiodicity exists in the
regions below the $\mu _{1}$ curve and above the $\mu _{2}$ curve. By
quasiperiodicity we mean that quasiperiodic motion is observed in the time
period $t\leq T=1\times 10^{7}.$ And we assume, (without proof) that the
actual critical couplings for the transitions are very near the plotted
curves of Fig. 3(a).

Both $\mu _{1}(N)$ and $\mu _{2}(N)$ curves have well behaved power law
scalings%
\begin{equation}
\mu _{1}(N)\propto \frac{1}{N^{\alpha }},\text{ \ }\alpha \approx 3.82
\end{equation}%
\begin{equation}
\mu _{2}(N)\propto N^{\beta },\text{ \ }\beta \approx 1.92
\end{equation}%
The solid lines $\mu _{1}(N)$ and $\mu _{2}(N)$ are given by $A/N^{\alpha },$
$A\approx 1\times 10^{-3},$ $\alpha \approx 3.82$ and $\mu _{2}(N)=BN^{\beta
},$ $B\approx 2.4,$ $\beta \approx 1.92,$ respectively. Equations (12)\ and
(13) are numerical results. We do not know how to derive the exponents $%
\alpha $ and $\beta $ analytically, and even have no idea at present to
explain them. We suggest these scaling may be model dependent and do not
have universal behavior. For instance, the scaling (13)\ is different from
the scaling Casetti and coworkers obtained for the FPU model\cite{revisite}.

With the initial condition (5) the critical point $\mu _{1}^{^{\prime }}$
and $\mu _{2}^{^{\prime }}$ are also related to $N$. The scalings are shown
in Fig. 3(b), which read 
\begin{eqnarray}
\mu _{1}^{^{\prime }}(N) &\propto &N^{\gamma }\text{ },\text{ \ }\gamma
\approx 0.57 \\
\mu _{2}^{^{\prime }}(N) &\propto &N^{\beta }\text{ },\text{ \ }\beta
\approx 1.95
\end{eqnarray}

\section{conclusion and discussion}

In conclusion, we have investigated the statistical behavior of the
classical $\phi ^{4}$ lattice from its microscopic dynamics. The particular
attention focuses on the transitions to chaos, equipartition behavior and
equilibrium state. By varying the coupling strength, two transitions to
equipartition from both small and large couplings are observed. The scaling
behaviors of these couplings against the system size are revealed
numerically.

However, there are still some problems unsolved. First, it is difficult to
precisely fix the transition points, because near the transitions the system
may show zero LE for extremely long period, and suddenly jump to positive LE
after a certain time. Thus, by a finite time simulation, one cannot
guarantee zero LE, and consequently cannot definitely determine the critical
conditions. In this paper, we apply an time truncation in identifying $\mu
_{1}$ and $\mu _{2}$, i.e., we regard the largest LE to be zero if its zero
value is kept for a given very long while still finite time [see Ref.\cite%
{revisite}]. We guess that the actual transition points are not far from the
critical couplings shown in Figs. 1-3. We will try a systematical method to
convincingly identify the transitions in the future work. Moreover, the
mechanism underlying the scaling relations (12), (13) and (15)\ are still
unclear, and worthwhile further investigating.

\begin{center}
{\bf ACKNOWLEDGEMENT}
\end{center}

This research was supported by the National Natural Science Foundation of
China, the Nonlinear Science Project of China, and the Foundation of
Doctoral training of Educational Bureau of China.

Captions of figures

Fig. 1 The largest LE of Eqs. (3) $\lambda _{1}$ for the fixed average
energy $\overline{E}=\frac{1}{N}\stackrel{N}{%
\mathrel{\mathop{\sum }\limits_{i=1}}%
}E_{i}=1.25\times 10^{-3}.$ (a) The largest LE vs the time $T$ with $T$
being the time for computing $\lambda _{1}.$ $\mu =6.9\times 10^{-7},$ $%
N=64, $ and the initial condition $(4)$ is taken. It is shown that $\lambda
_{1}(T) $ remains zero for a long period, and it bursts to be positive after
a certain time. (b) the same as (a) with $\mu =2.5$. (c) $\lambda _{1}$ vs $%
\mu $ for $N=32.$ Circles are for the initial condition (4)\ and triangles
are for the initial condition (5). The transition points $\mu _{1}$, $\mu
_{2}$ and $\mu _{1}^{^{\prime }}$, $\mu _{2}^{^{\prime }}$ are determined by
computations of $T=1\times 10^{7}.$ (d)\ The same as (c)\ with $N=64.$

Fig. 2 $N=64.$ $\eta _{m}$ [(a)] and $\eta _{s}$ [(b)] defined in Eq. (11)\
plotted vs $\mu .$ In (a)\ the initial condition (4)\ is used and in (b)\
Eq. (5)\ is used. $\eta _{m}$ has discontinuities at two transition points $%
\mu _{1}$ and $\mu _{2}$ of Fig. 1(d), and $\eta _{s}$ jumps down at the
transition point $\mu _{1}^{^{\prime }}$ of Fig. 1(d). (c), (d)\ $\mu =0.1.$
The $x_{i}$ [(c)] and $p_{i}$ [(d)] probability distributions of the sites $%
i=1,2,....N$ for both initial conditions (4) and (5).

Fig. 3 (a) The transition couplings $\mu _{1}$ and $\mu _{2}$ plotted for
different systems size with the initial condition (4) being taken. In the
region between $\mu _{1}$ and $\mu _{2}$ lines, the systems motion is
chaotic, and equipartition behavior can be identified. $\mu _{1}$ and $\mu
_{2}$ have power law relations with the size $N$ as $\mu _{1}$ $\propto
1/N^{3.82},$ $\mu _{2}\propto N^{1.92}.$ The exponents are measured by the
slopes of the two solid lines. (b) The same as (a) with the initial
condition (5) used. $\mu _{1}^{^{\prime }}$ $\propto N^{0.57},$ $\mu
_{2}^{^{\prime }}\propto N^{1.95}.$

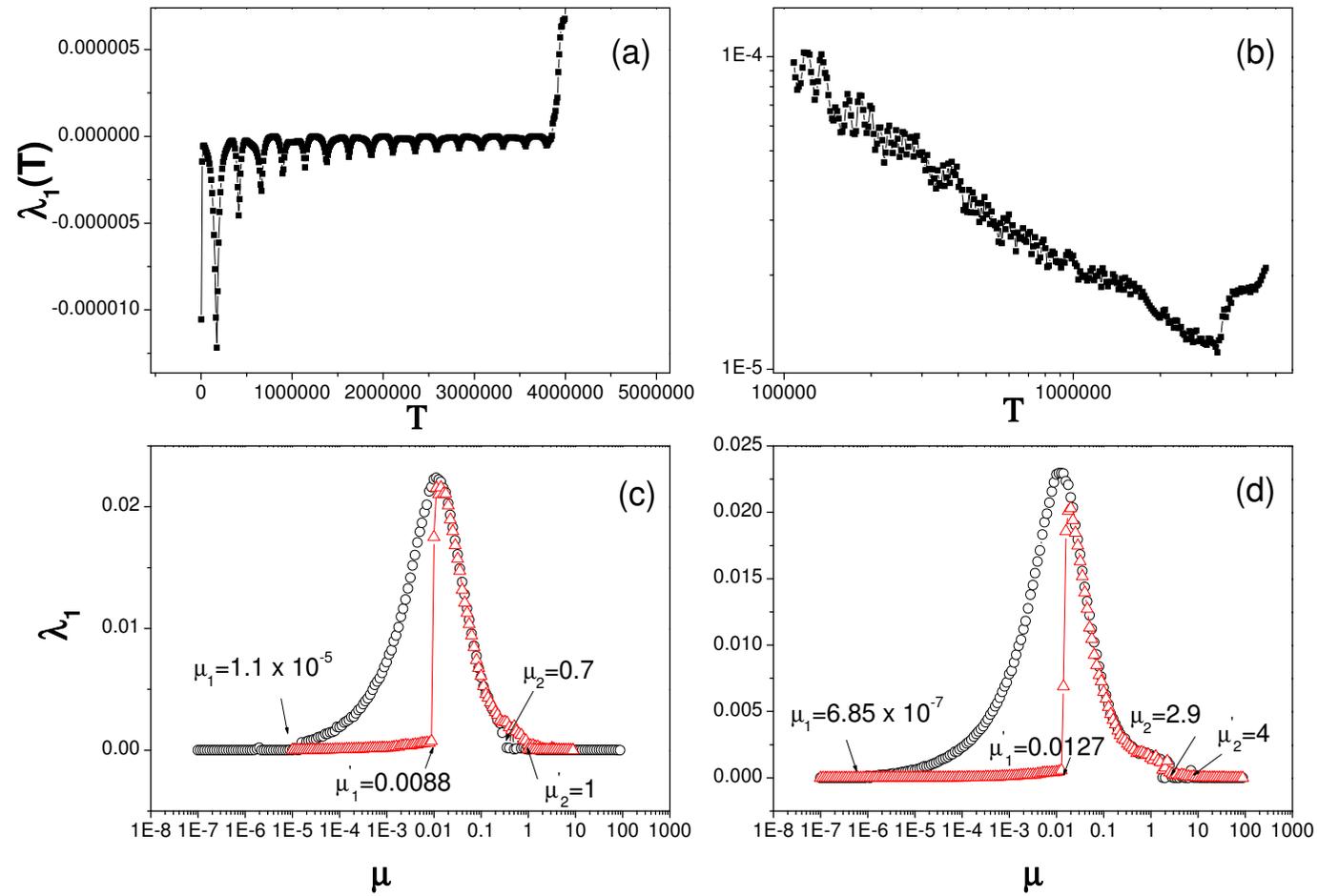
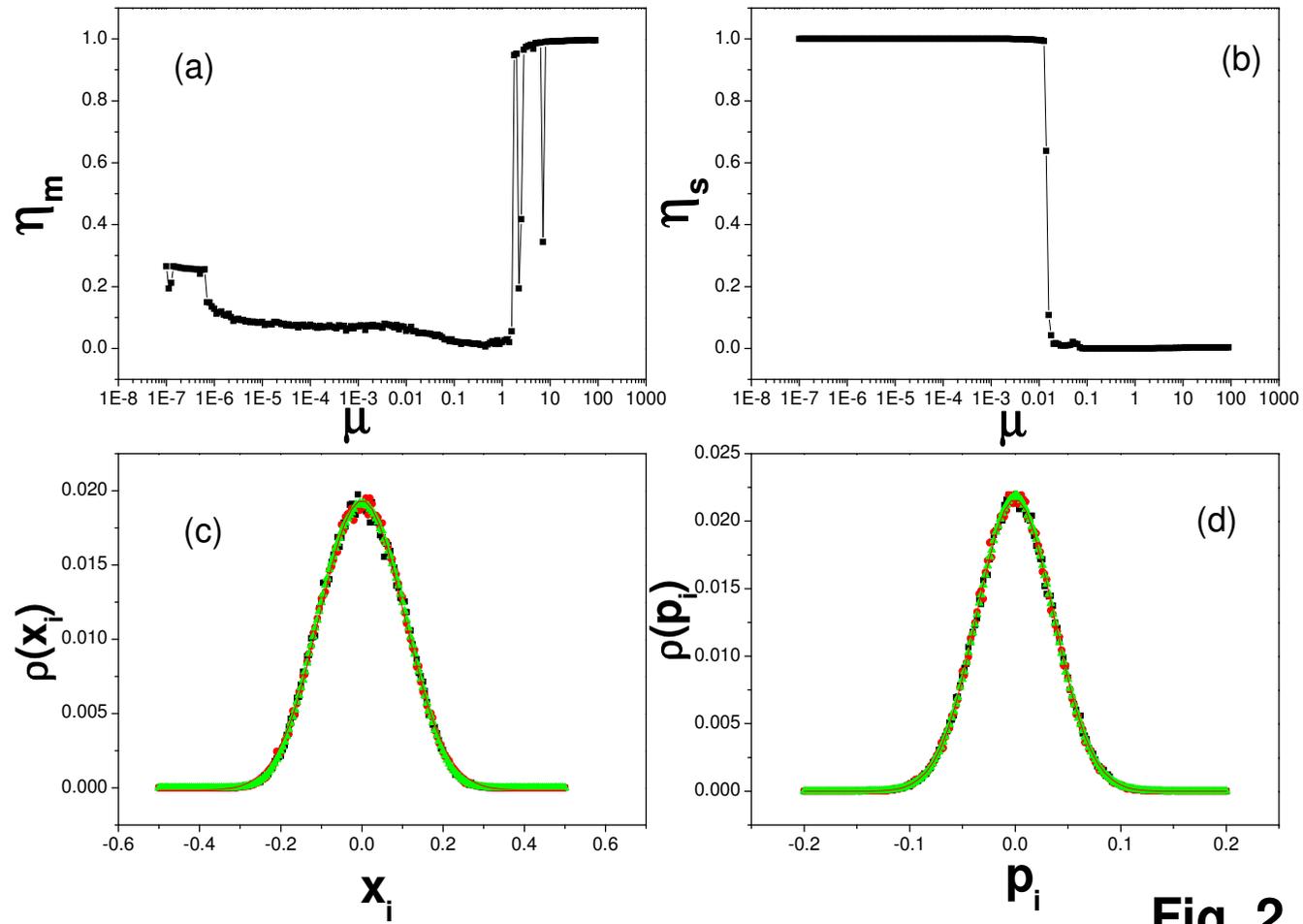
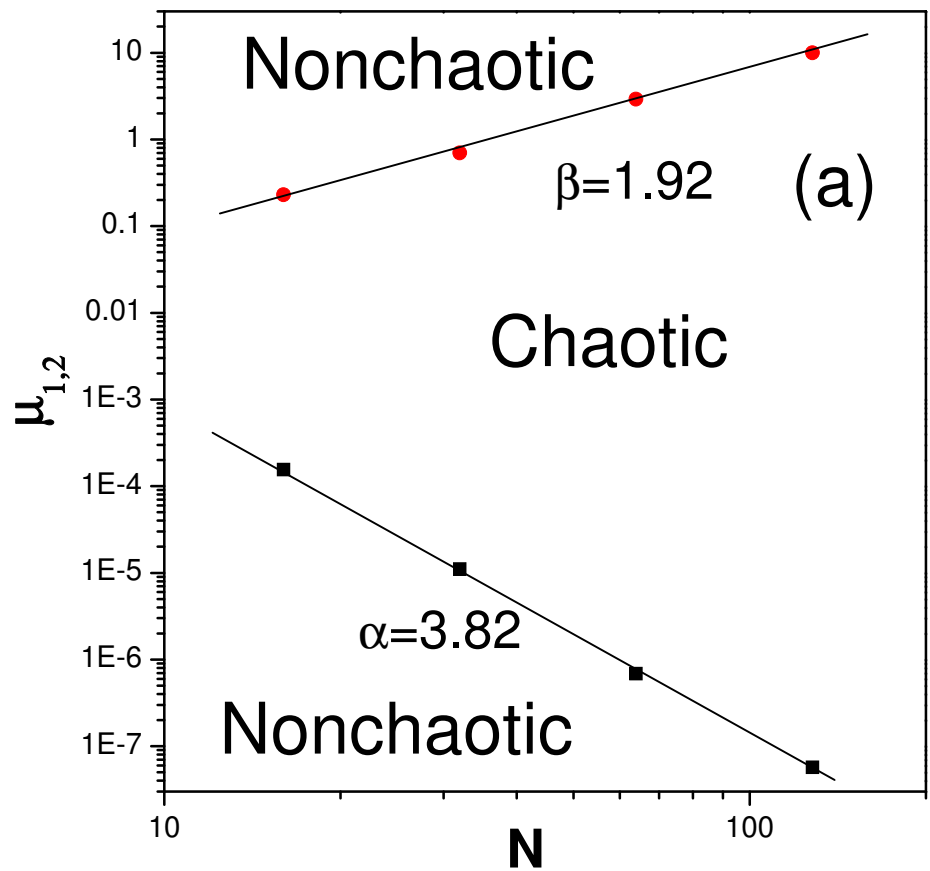
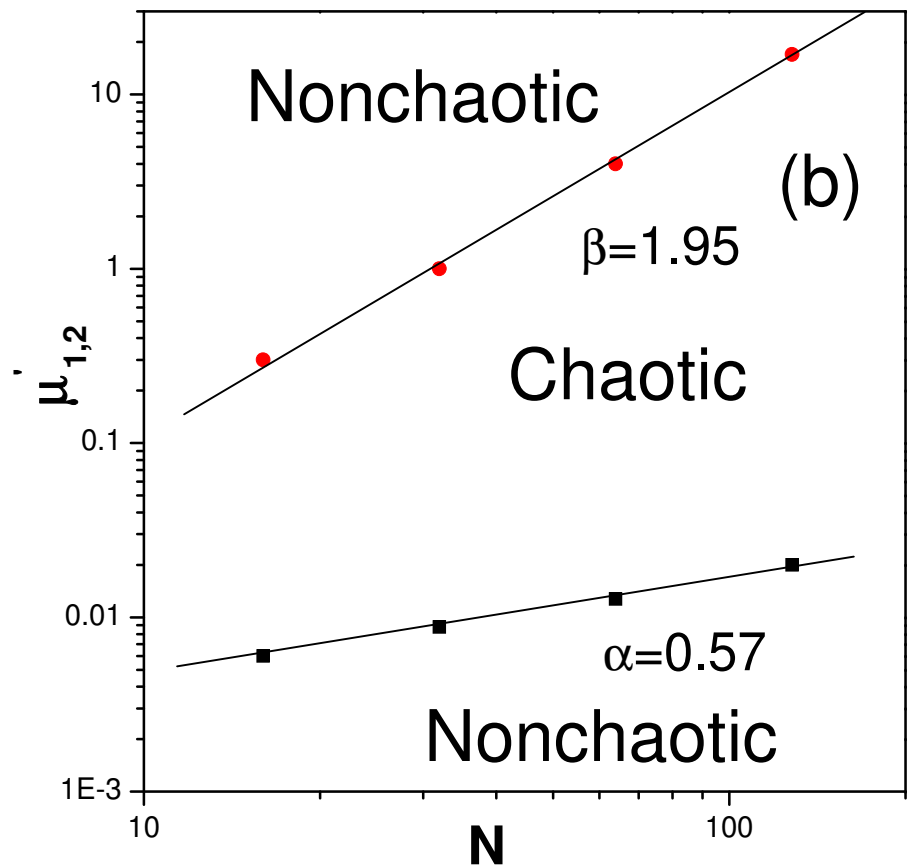
\end{document}